\documentclass[12pt]{article}
\usepackage{geometry}
\usepackage[utf8]{inputenc}
\usepackage{enumitem,amssymb}
\usepackage{ragged2e}
\newlist{thematic}{itemize}{8}
\setlist[thematic]{label=$\square$}
\usepackage{pifont}
\usepackage{microtype}
\usepackage[raggedright]{titlesec}
\titlespacing*{\section}{0pt}{\baselineskip}{0.5\baselineskip}
\titlespacing*{\subsection}{0pt}{0.5\baselineskip}{0.25\baselineskip}
\usepackage{graphicx}
\usepackage[usenames,dvipsnames,svgnames]{xcolor}
\definecolor{DeepBlue}{rgb}{0.1, 0.25, 0.55}
\definecolor{DeepGreen}{rgb}{0.0, 0.25, 0.0}
\usepackage[colorlinks,
            citecolor=DeepBlue,
            urlcolor=DeepGreen,
            linkcolor=DeepBlue]{hyperref}
\usepackage{siunitx}
\usepackage{subcaption}
\usepackage[p]{newtxtext}
\usepackage{newtxmath}
\usepackage[varl]{zi4}
\usepackage[autocite=superscript,
            autopunct=true,
            backend=biber,
            citestyle=verbose,
            eprint=true,
            style=phys]{biblatex}
\makeatletter
\newcommand\footref[1]{\protected@xdef\@thefnmark{\ref{#1}}\@footnotemark}
\makeatother

\newcommand{\ie}{\emph{i.e.}}
\newcommand{\eg}{\emph{e.g.}}
\newcommand{\tss}{\textsuperscript}

\DeclareSourcemap{
  \maps[datatype=bibtex]{
    \map{
      \pertype{article}
      \step[fieldsource=journal, final]
      \step[fieldset=archivePrefix, null]
      \step[fieldset=eprint, null]
      }
    }
  }
 
\newcommand{\iligoBnsHorizonZ}{0.01}
\newcommand{\iligoBbhThirtyHorizonZ}{0.116}

\newcommand{\otwoBnsHorizonZ}{0.047}
\newcommand{\otwoBbhThirtyHorizonZ}{0.56}

\newcommand{\aplusBnsHorizonZ}{0.184}
\newcommand{\aplusBbhThirtyHorizonZ}{2.7}

\begin{document}
{\raggedright
\huge
Astro2020 Science White Paper \linebreak

The US Program in Ground-Based Gravitational Wave Science: Contribution from the LIGO Laboratory \linebreak

\normalsize


\noindent
\begin{tabbing}
\textbf{Thematic Areas:} \hspace*{60pt} \= $\square$ Planetary Systems \hspace*{15pt} \= $\square$ Star and Planet Formation \\
$\blacksquare$ Formation and Evolution of Compact Objects \>\> $\blacksquare$ Cosmology and Fundamental Physics \\
$\square$ Stars and Stellar Evolution \> $\square$ Resolved Stellar Populations and their Environments \\
$\square$ Galaxy Evolution \> $\blacksquare$ Multi-Messenger Astronomy and Astrophysics \linebreak
\end{tabbing}

\textbf{Principal Author:}

Name: David Reitze
 \linebreak						
Institution: LIGO Laboratory, California Institute of Technology
 \linebreak
Email: dreitze@caltech.edu
 \linebreak
Phone: +1--626--395--6274  
 \linebreak
 
\textbf{Co-authors:} \\
LIGO Laboratory, California Institute of Technology, Pasadena, California 91125, USA \\
LIGO Laboratory, Massachusetts Institute of Technology, Cambridge, Massachusetts 02139, USA \\
LIGO Hanford Observatory, Richland, Washington 99352, USA \\
LIGO Livingston Observatory, Livingston, Louisiana 70754, USA \linebreak

}

\noindent
\textbf{Abstract:}
Recent gravitational-wave observations from the LIGO and Virgo observatories have brought a sense of great excitement to scientists and citizens the world over. Since September 2015, 10 binary black hole coalescences and one binary neutron star coalescence have been observed. They have provided remarkable, revolutionary insight into the ``gravitational Universe'' and have greatly extended the field of multi-messenger astronomy. At present, Advanced LIGO can see binary black hole coalescences out to redshift \num[round-mode=figures,round-precision=1]{\otwoBbhThirtyHorizonZ{}} and binary neutron star coalescences to redshift \num[round-mode=figures,round-precision=1]{\otwoBnsHorizonZ{}}.
This probes only a very small fraction of the volume of the observable Universe. However, current technologies can be extended to construct ``3\tss{rd} Generation'' (3G) gravitational-wave observatories that would extend our reach to the very edge of the observable Universe. The event rates over such a large volume would be in the hundreds of thousands per year (\ie{} tens per hour). Such 3G detectors would have a 10-fold improvement in strain sensitivity over the current generation of instruments, yielding signal-to-noise ratios of 1000 for events like those already seen. Several concepts are being studied for which engineering studies and reliable cost estimates will be developed in the next 5 years.

\pagebreak

\section{Introduction}
\label{sec:intro}

Long-baseline laser interferometry~\autocite{Bond2017,Adhikari:2013kya} has given life to gravitational-wave astronomy and expanded the vision of multi-messenger observation. When the current generation of interferometric detectors are brought to their full strain sensitivities, they will see binary neutron star coalescences out to redshift $z \simeq \num[round-mode=figures,round-precision=1]{\aplusBnsHorizonZ{}}$ and binary black hole coalescences out to $z \simeq \num[round-mode=figures,round-precision=1]{\aplusBbhThirtyHorizonZ{}}$~\autocite{T1800042}, with nearby events having signal-to-noise ratios of order 100. With a future generation of more sensitive detectors in new, larger facilities, it will be possible to see these binary systems out to redshifts $z > 10$, with nearby events having signal-to-noise ratios in excess of 1000. 
This paper presents a synopsis of the scientific opportunities afforded by 3G detectors; it also outlines needed technical developments to enable them.


\section{A brief history of the field}
\label{sec:history}

LIGO consists of two NSF-funded~\autocite{NSF} US facilities that host \SI{4}{km} long, interferometric gravitational-wave detectors, one in Livingston, Louisiana, and the other in Hanford, Washington. The facilities were constructed from 1992 to 1999, and the initial LIGO detectors (iLIGO) made observations from 2002 to 2010, ultimately reaching a sensitivity to neutron stars out to redshift $z \simeq \num[round-mode=figures,round-precision=1]{\iligoBnsHorizonZ{}}$ and black holes to redshift $z \simeq \num[round-mode=figures,round-precision=1]{\iligoBbhThirtyHorizonZ{}}$---though no detections were made~\autocite{Abbott:2007kv,LIGO:2012aa}.

The second-generation detectors, known as Advanced LIGO (aLIGO), were funded~\autocite{NSF,MPS,STFC,ARC} starting in 2008 and installed at both sites from 2011 to 2014. aLIGO was designed to achieve a 10-fold increase in  sensitivity over iLIGO---providing a 1000-fold increase in sensitive volume---by reducing the instrument's high-frequency noise floor and extending the sensitive band to lower frequency~\autocite{TheLIGOScientific:2014jea}. Observations began in September 2015 with about four times the sensitivity of iLIGO~\autocite{TheLIGOScientific:2016agk}. On September~14, 2015, the first gravitational waves were detected: a chirp signal (GW150914) from the merger of two black holes \SI{410}{Mpc} from Earth~\autocite{Abbott:2016blz}. Since then, aLIGO's sensitivity has been improved and more observations have been made.

The first two aLIGO observing runs (O1, from Sep~2015 to Jan~2016, and O2, from Nov~2016 to Aug~2017) totaled a calendar year of observing time. Near the end of O2, the Virgo observatory (a \SI{3}{km} detector in Cascina, Italy)~\autocite{TheVirgo:2014hva} joined LIGO, and critically contributed to the first observation of a binary neutron star merger (GW170817) by reducing the sky location area from a few hundred to a few tens of square degrees~\autocite{TheLIGOScientific:2017qsa}. This enabled 
approximately 90 optical and radio astronomy telescopes to identify and study electromagnetic counterparts to the gravitational waves~\autocite{GBM:2017lvd}. In O1 and O2 combined, 10 black hole coalescences and one neutron star collision were observed~\autocite{LIGOScientific:2018mvr}. Observing run O3 will begin in April 2019 and run for about one year. KAGRA, an underground \SI{3}{km} detector in Japan~\autocite{Akutsu:2019rba}, 
is planning to join O3 toward the end of 2019 at a reduced sensitivity. By the middle of the next decade, LIGO will install and commission an upgrade (referred to as ``A+'') to significantly improve the aLIGO detectors beyond their design sensitivity target (by a factor of 5 in detection rate for compact binary mergers compared to aLIGO at design sensitivity).  Additionally, a third LIGO detector will be brought online in a new \SI{4}{km} facility in India.

\section{Science case for an international network of 3G detectors}
\label{sec:science}

The worldwide gravitational-wave, high energy astrophysics, and nuclear physics communities are highly energized by the recent discoveries and the prospects for an extremely rich science program ahead. The Gravitational Wave International Committee (GWIC)~\autocite{GWIC} is developing a set of science white papers detailing 
3G gravitational-wave astronomy based on a 3G gravitational-wave detector network consisting of ``Cosmic Explorer''~(CE)~\autocite{CE,Evans:2016mbw} and ``Einstein Telescope''~(ET)~\autocite{ET,Punturo:2010zz} detectors, as described in the next section. Here we summarize the science case for Cosmic Explorer, which delivers two significant advances over the current generation of gravitational-wave detectors: (a)~nearby sources will be detected with very high signal-to-noise ratio (SNR); and (b)~sources will be detectable to redshifts greater than 10 (see Fig.~\ref{fig:strain_horizon})~\autocite{Vitale:2016aso}, allowing studies of their evolution and providing access to the first stars in the Universe. 

\subsection{Black holes}

\begin{figure}[t]
    \centering
    \begin{subfigure}{0.525\textwidth}
        \includegraphics[width=\textwidth]{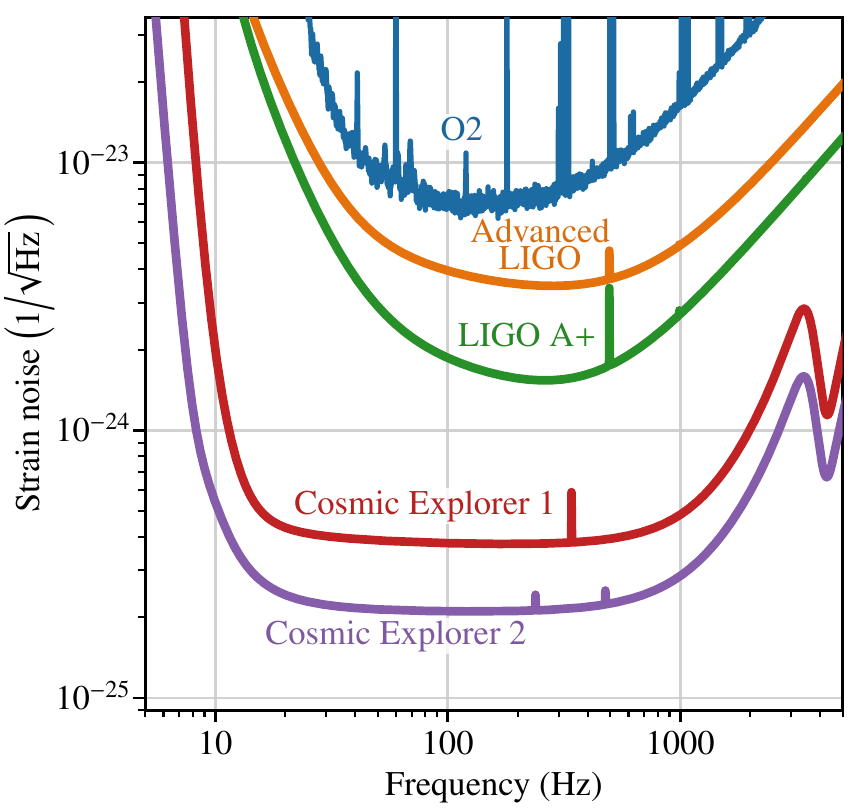}
        \label{fig:strain}
    \end{subfigure}%
    \begin{subfigure}{0.475\textwidth}
        \includegraphics[width=\textwidth]{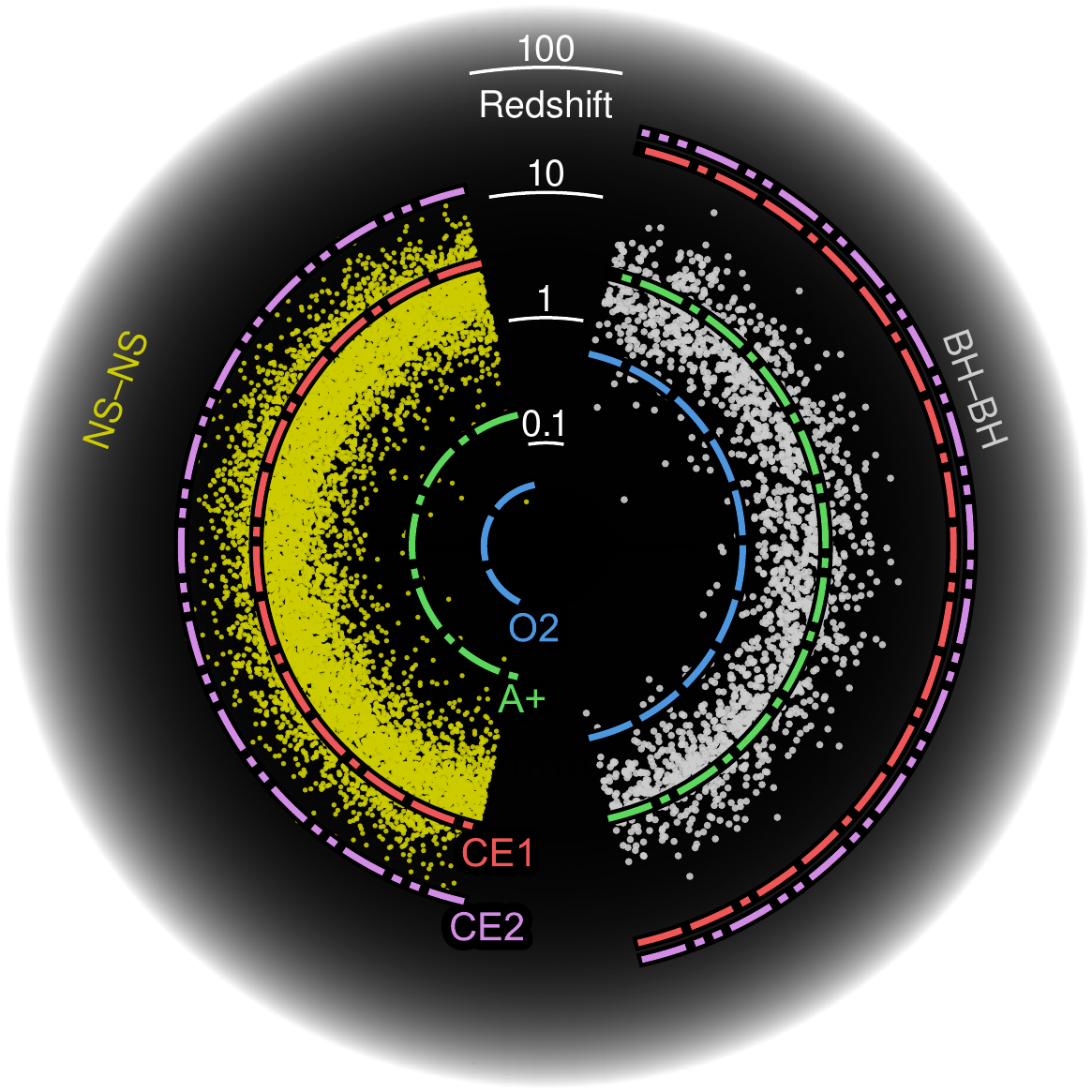}
        \label{fig:horizon}
    \end{subfigure}
    \caption{\emph{Left side:} Cosmic Explorer projected strain noise for Stage~1 (during the 2030s) and Stage~2 (2040s), compared with the strain noise achieved by Advanced LIGO during observing run O2, as well as designed noise performance for Advanced LIGO, and LIGO~A+. Less strain noise indicates better strain sensitivity. \emph{Right side:} Astrophysical response distance~\autocite{Chen:2017wpg} of Advanced LIGO at O2 sensitivity, LIGO~A+, and Cosmic Explorer (Stages~1 and~2), plotted on top of a population of 1.4--1.4$M_\odot$ neutron star mergers (yellow) and 30--30$M_\odot$ black hole mergers (gray), assuming a Madau--Dickinson star formation rate~\autocite{Madau:2014bja} and a typical merger time of \SI{100}{Myr}. The radial distribution of points accounts for the detector-frame merger rate per unit redshift.
}
    \label{fig:strain_horizon}
\end{figure}

\textbf{Physics:} 3G detectors will enable unprecedented tests of the strong-field dynamics of gravity.  Some specific effects include the ringdown of a black hole merger~\autocite{Gossan:2011ha,Berti:2016lat}, detailed measurements of spin-orbit gravitational interactions~\autocite{Kidder:1992fr,PhysRevD.49.6274,Vitale:2016icu}, the non-linear memory effects in the metric that accompany gravitational radiation~\autocite{Christodoulou:1991cr,Yang:2018ceq}, and direct tests of the Hawking area theorem~\autocite{Cabero:2017avf}. Many of these tests rely on obtaining very high SNR signals, which is possible with the superior noise performance of the 3G detectors (Fig.~\ref{fig:strain_horizon}, left side).

\vspace{0.175\baselineskip}\noindent%
\textbf{Astronomy:} The right side of Fig.~\ref{fig:strain_horizon} shows how Cosmic Explorer will be able to detect black hole mergers throughout cosmic history, thus exploring both the importance of different formation channels (field formation versus dynamical capture) and how these evolved with time. Evidence for black holes from population~III stars might help in understanding the formation of supermassive black holes and the galaxies they seed~\autocite{Sesana:2009wg}. The mass and spin distribution of black holes will reveal their origin and key properties of their progenitors, such as the magnitude of supernovae kicks~\autocite{OShaughnessy:2017eks}.
If primordial stellar-mass black hole binaries exist, Cosmic Explorer will detect them if they merge at $z \lesssim 100$. A network of Cosmic-Explorer-class detectors will provide high-SNR signals from all epochs of stellar-mass binary black hole collisions in the Universe.

\subsection{Neutron stars}

\textbf{Physics:} Capturing high-SNR binary neutron star coalescences is the most promising way of measuring the equation of state of nuclear matter through precise measurements of the phase evolution of the gravitational-wave signal~\autocite{Flanagan:2007ix}. More detailed information about the composition of neutron stars may be available from the gravitational-wave emission from the post-merger collision remnant~\autocite{Stergioulas:2011gd}. Furthermore, measurements of the masses and spins of the component neutron stars, and of the remnant, will lead to a better understanding of the the yield of heavy metals produced by these systems~\autocite{Metzger:2016pju}, and the maximum mass of neutron stars\autocite{Margalit:2017dij}. As shown in Fig.~\ref{fig:strain_horizon} (right side), Advanced LIGO and A+ will only detect a very small fraction of binary neutron star coalescences; 
Cosmic Explorer will have access to the entire coalescing population.

Gravitational waves can also be produced by isolated spinning neutron stars, as long as they are not perfectly spherical~\autocite{Zimmermann:1979ip}. Such gravitational waves have not yet been detected, and the amplitude of emission from these sources is not known. Such a signal is expected to be continuous and nearly monochromatic, and would provide information about neutron star ellipticity, moment of inertia, and again nuclear equation of state.

\vspace{0.175\baselineskip}\noindent%
\textbf{Astronomy:} 3G detectors can detect binary neutron star coalescences to redshifts of unity and above~\autocite{Mills:2017urp}. Many of these events will be accompanied by detectable electromagnetic emission, as in GW170817. With hundreds of thousands of sources per year, 3G detectors will enable precise measurements of the mass function of neutron stars and their merger rate. Routine detections of electromagnetic counterparts will allow us to more fully understand the details of the electromagnetic emission, including the nuclear-decay-powered kilonova~\autocite{Metzger:2016pju}. Localizing the hosts will also provide information about the galaxies in which the systems were formed, such as their age and metallicity~\autocite{TheLIGOScientific:2016htt}.

\vspace{0.175\baselineskip}\noindent%
\textbf{Cosmology:}
Gravitational waves provide an independent way to measure cosmological parameters. Each signal provides a direct measurement of the luminosity distance to the source. If the source redshift can be obtained by other means (\eg{} an electromagnetic counterpart, as for GW170817~\autocite{Abbott:2017xzu}, or observation of tidal effects in the waveform~\autocite{Messenger:2011gi}), one can then solve for the cosmological parameters. Advanced LIGO will continue to detect binary neutron star coalescences in the local universe, and hence will improve constraints on the Hubble constant~\autocite{Chen:2017rfc}; it will not, however, have access to other fundamental quantities, such as $\Omegaup_\Lambdaup$, $\Omegaup_\text{m}$, or the equation of state of dark energy. Cosmic Explorer opens the door to the measurement of the full set of cosmological parameters. Furthermore, it will reveal hundreds of thousands of sources per year up to high redshifts, vastly improving the precision of cosmological parameters and providing insights on the isotropy of the universe at different epochs.

\subsection{Other science targets}

3G detectors will improve the prospects for the detection of gravitational waves from core-collapse supernovae~\autocite{Powell:2018isq}, and a 3G network might detect a stochastic gravitational-wave relic from the early Universe~\autocite{Regimbau:2016ike}. Additionally, 3G detectors may reveal gravitational-wave sources that are entirely unanticipated.

\section{Pathways for future development of the field}
\label{sec:future}





\subsection{A global 3G network and the US vision for 2020--2030 and beyond}

3G detector concepts generally aim at achieving a factor of 10 increase in strain sensitivity compared to aLIGO and Advanced Virgo. To enable multi-messenger astronomy and high-precision science, it is critically important to determine the position of gravitational-wave sources on the sky. This requires a network of 3 or more 3G detectors with site locations around the world~\autocite{Mills:2017urp}.

The NSF is funding a study of ways in which the US can contribute to an international 3G effort. In this study, coordinated with GWIC, various detector concepts will be evaluated and costed at a preliminary level.
Cosmic Explorer~\autocite{CE} is a concept for a new facility with much longer arm lengths to boost sensitivity (\eg{} \SI{40}{\km} 
over the the current LIGO \SI{4}{\km} arm lengths).  At present, two Cosmic Explorer detectors, with locations to be determined, are envisioned as part of an international network that would include the European Einstein Telescope~\autocite{ET} as a third observatory. Second-generation detectors will play a supporting role in the global network as the first 3G detectors come online.

\subsection{Cosmic Explorer: a 40\texorpdfstring{\,}{ }km detector}

\begin{figure}[t]
    \centering
    \includegraphics[width=\textwidth]{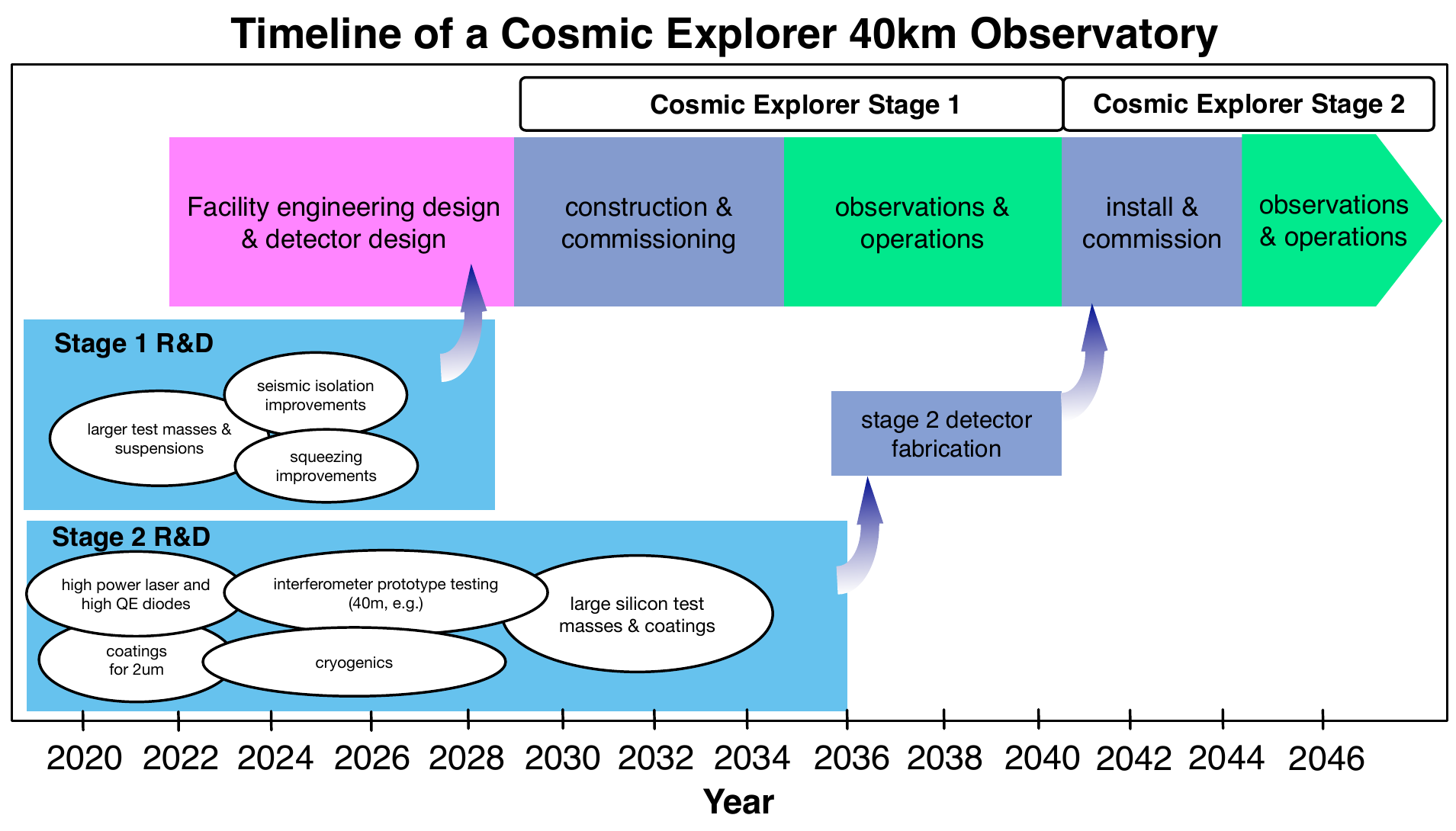}
    \caption{A plausible timeline for the development of Cosmic Explorer, with major milestones for funding and for the R\&D program to support the CE Stage~2 upgrade to new technologies. Periods of observations are represented in green, while periods of fabrication and of down-time of the instruments for installation and commissioning are shown in gray.}
    \label{fig:timeline}
\end{figure}

Cosmic Explorer will use longer arm lengths and new technologies to achieve a factor of 10 or more strain sensitivity improvement over the current generation of gravitational-wave detectors~\autocite{Evans:2016mbw}. As with LIGO, we envision a staged program for Cosmic Explorer (CE):
\begin{itemize}[nosep,itemsep=0.2\baselineskip,topsep=0.2\baselineskip]
    \item \textbf{CE Stage 1} (CE1, targeting operations in the late 2030s) relies on extensions of technologies demonstrated and being developed for A+.
    \item \textbf{CE Stage 2} (CE2, starting operations in the mid-2040s) consists of a major upgrade of CE that will exploit the full potential of the new facility by reducing thermal noise through the use of cryogenics and new materials for test masses and coatings. A promising route currently under investigation consists of employing silicon test masses and amorphous silicon coatings operating at \SI{123}{\kelvin}, with \num{1.5} or \SI{2}{\um} laser light; these are major changes from existing technologies and will require long-term R\&D to develop.
\end{itemize}
This tentative timeline is captured by Fig.~\ref{fig:timeline}, as well as elements of the R\&D program necessary to support the construction of the detector and its upgrade to new lasers and new mirror materials.

Fig.~\ref{fig:strain_horizon} (left side) shows the sensitivity progression of the detectors, from the most recent Advanced LIGO observing run (O2) to CE2. Improving the strain sensitivity beyond what can be done in existing facilities is critical to delivering the science goals discussed above. Running second-generation detectors for a longer period of time would increase linearly the number of detected sources; however, these sources would still lie in the local universe and have low to moderate SNRs. On the other hand, increasing the strain sensitivity extends the reach of the detectors to a redshift of many. This gives access to remote sources that cannot possibly be detected with second-generation detectors, while also measuring nearby sources with SNRs of thousands. Cosmic Explorer will not only collect a much greater number of signals compared to Advanced LIGO; it will also shed light on a gravitational-wave universe that is currently inaccessible to us. 

Beside requiring new facilities, 
Cosmic Explorer will 
require a series of technical developments in several areas. These include larger test masses, suspension systems able to sustain heavier masses, improved vibration isolation, improved angular alignment control, introduction of greater optical power, higher levels of frequency-dependent squeezing, and mitigation of gravity gradients. A significant issue is the reduction of thermal noise in the mirror coatings, which may require multiple approaches, including changes in laser wavelength, new mirror materials, and cryogenic operation. An active research program in the next decade will define the best strategy to maximize the scientific output of Cosmic Explorer. Engineering studies will be needed to establish designs and formal cost estimates, and upgrades to laboratory prototyping facilities are required to carry out the necessary R\&D.

\section{Outlook}
\label{sec:funding}

The construction cost of the envisioned 3G detectors is not well known, as full-fledged designs for 3G detectors are not yet available. To move forward we will request funding for engineering and costing studies, and funds to support a basic research program with identified technical goals that must be reached to finalize the design.

\pagebreak

\printbibliography{}

\end{document}